\documentstyle[prl,aps,multicol,epsfig]{revtex}

\input epsf.tex

\newcommand{\ba}{\begin{eqnarray}}
\newcommand{\ea}{\end{eqnarray}}
\newcommand{\ban}{\begin{eqnarray*}}
\newcommand{\ean}{\end{eqnarray*}}
\newcommand{\ket}[1]{\left|#1\right\rangle}
\newcommand{\bra}[1]{\mbox{$ \langle #1 | $}}
\newcommand{\moy}[1]{\langle #1 \rangle}

\begin{document}

\title{Nonlocality of cluster states of qubits}
\author{Valerio Scarani$^{1}$, Antonio Ac\'{\i}n$^{2}$, Emmanuel Schenck$^{3,4}$, Markus Aspelmeyer$^{3}$ }
\address{
$^{1}$ Group of Applied Physics, University of Geneva, 20, rue de
l'Ecole-de-M\'edecine, CH-1211 Geneva 4, Switzerland\\
$^{2}$ ICFO-Institut de Ci\`encies Fot\`oniques, 29 Jordi Girona, 08034 Barcelona, Spain.\\
$^{3}$ Institut f\"ur Experimentalphysik, Universit\"at Wien,
Boltzmanngasse 5, A-1090 Vienna, Austria\\
$^{4}$ Permanent address: Ecole normale sup\'erieure, 45, rue
d'Ulm, 75005 Paris, France}
\date{\today}
\maketitle \maketitle

\begin{abstract}
We investigate cluster states of qubits with respect to their
non-local properties. We demonstrate that a
Greenberger-Horne-Zeilinger (GHZ) argument holds for any cluster
state: more precisely, it holds for any partial, thence {\em
mixed}, state of a small number of connected qubits (five, in the
case of one-dimensional lattices). In addition, we derive a new
Bell inequality that is maximally violated by the 4-qubit cluster
state and is not violated by the 4-qubit GHZ state.
\end{abstract}

\begin{multicols}{2}

\section{Introduction}

In its most widespread image, a quantum computer is depicted by an
array of initially uncorrelated qubits that pass through a network
of logic gates in which they become entangled \cite{nc}. In 2001,
Raussendorf and Briegel noticed that one can adopt a different
philosophy and described the so-called {\em one-way quantum
computer} \cite{br2}. In this view, the entanglement is
distributed once for all by preparing a peculiar entangled state
of all the qubits; the logic gates are then applied as sequences
of only single-qubit measurements. Remarkable entanglement
properties are needed to achieve this computational power: in
particular, Greenberger-Horne-Zeilinger (GHZ) states, though in
some sense maximally entangled, lack this power, and indeed can be
simulated by a polynomial amount of communication \cite{tessier}.
Suitable $N$-qubit states for universal, scalable quantum
computation are the so-called {\em cluster states} \cite{br1}.
Motivated by this discovery, many works have been devoted in the
last few years to the properties of those states \cite{others},
links have been found in particular with error-correction theory
\cite{error}. In this paper we study the cluster states under the
perspective of their non-locality properties \cite{div}.

A brief review of the definition and main properties of the
cluster states, following \cite{br2,br1}, is a necessary
introduction. For convenience, through all this paper we adopt the
notations $X=\sigma_x$, $Y=\sigma_y$ and $Z=\sigma_z$ for the
Pauli matrices. A {\em cluster} is a $N$-sites $d$-dimensional
square lattice with connections among the sites, that define a
notion of neighborhood. For any site $a$ of the lattice, one
defines the operator \ba S_a &=& X_a\,\bigotimes_{b\in
neigh(a)}Z_b \label{sa}\ea where $neigh(a)$ is the set of all the
neighbors of $a$. The operators $\{S_a, a\in \mbox{lattice}\}$
form a complete family of commuting operators on the lattice; a
{\em cluster state} is any of their common eigenvectors. We note
here that this construction can actually be done for any graph,
not only for square lattices, leading to the notion of {\em graph
states}; but in this paper, we stick to cluster states and will
discuss possible extensions to all graph states only in the final
Section. For definiteness, we consider the cluster state
$\ket{\Phi_N}$ associated with all eigenvalues being $+1$, that
is, determined by the family of equations \ba S_a
\ket{\Phi_N}\,=\,\ket{\Phi_N} &&\mbox{ for all
$a$}\,.\label{family} \ea We start by considering cluster states
built on a one-dimensional lattice \cite{note2}, that we note
$\ket{\phi_N}$. For such a lattice, $\ket{\phi_2}$ and
$\ket{\phi_3}$ are locally equivalent to a maximally entangled
Bell state and to a GHZ state, respectively; the non-locality of
both has been thoroughly studied. The 4-qubit cluster state reads
\ba \ket{\phi_4}&=& \frac{1}{2}\,\ket{+}\ket{0}\ket{+}\ket{0}\,
+\, \frac{1}{2}\, \ket{+}\ket{0}\ket{-}\ket{1}\nonumber\\&&
+\,\frac{1}{2}\, \ket{-}\ket{1}\ket{-}\ket{0} \,+\, \frac{1}{2}\,
\ket{-}\ket{1}\ket{+}\ket{1} \ea where the one-qubit states are
defined as usual as $Z\ket{0}= \ket{0}$, $Z\ket{1}= -\ket{1}$ and
$X\ket{\pm}= \pm\ket{\pm}$. Note that the state $\ket{\phi_4}$ is
{\em not} locally equivalent to the 4-qubit GHZ state
$\ket{GHZ_4}=\frac{1}{\sqrt{2}} \big(\ket{0000}+\ket{1111}\big)$,
but to $\frac{1}{2} \big(\ket{0000}+\ket{0011}+
\ket{1100}-\ket{1111}\big)$.

The plan of the paper is as follows. Section \ref{sec4q} is
devoted to the the non-locality properties of $\ket{\phi_4}$: we
identify a GHZ argument for non-locality \cite{ghz}, from which in
turn a Bell inequality can be derived. The advantage of this
particular construction is that the inequality is optimized for
the state $\ket{\phi_4}$: it acts as witness discriminating
between $\ket{\phi_4}$, that violates it up to the algebraic
limit, and $\ket{GHZ_4}$, that does not violate it at all. In
Section \ref{secN1} we generalize the GHZ argument to the
$N$-qubit case for one-dimensional lattices, then in Section
{\ref{secNd}} for $d$-dimensional lattices. In both cases,
contrary to what happens for GHZ states, a GHZ argument for
non-locality can be found for the partial (thence {\em mixed})
states defined on small sets of connected qubits once all the
others are traced out. The result is quite surprising, since it
was commonly believed that the purity of quantum states was a
necessary condition for all-or-nothing violations of local
realism. Finally, in Section \ref{secgraph} we consider the larger
family of graph states.

\section{Non-locality of the 4-qubit state $\ket{\phi_4}$ on a 1-dimensional cluster}
\label{sec4q}

\subsection{GHZ argument}
\label{secghz4}

The 4-qubit cluster state $\ket{\phi_4}$ is defined by
(\ref{family}) \ba
\begin{array}{ccl} XZII=+1&&(E_1)
\\ ZXZI=+1&&(E_2)\\
IZXZ =+1&&(E_3)\\ IIZX =+1&&(E_4) \,.\end{array} \label{c4}\ea
These notations are shortcuts for $XZII\ket{\phi_4}= \ket{\phi_4}$
etc. Eleven similar equations can be obtained by multiplication
using the algebra of Pauli matrices: \ba
(E_1)\times(E_3)&\;:\;&XIXZ=+1 \label{ghz1} \\
(E_2)\times(E_3)&\;:\;&ZYYZ =+1 \label{ghz2}\\
(E_1)\times(E_3)\times(E_4)&\;:\;& XIYY =+1\label{ghz3}\\
(E_2)\times(E_3)\times(E_4)&\;:\;& ZYXY =-1\label{ghz4} \ea and
seven others which we won't use explicitly in this paper, namely
$YYZI=XZZX=ZXIX=IZYY=YYIX=YXXY=+1$ and $YXYZ=-1$. The fifteen
properties can be read directly from $\ket{\phi_4}\bra{\phi_4}$;
they are associated to the operators that, together with the
identity, form the Abelian group generated by $\{S_1,...,S_4\}$,
called {\em stabilizer group}. Note how a minus sign arises from
the multiplication of three consecutive equations, since in the
common site one has $ZXZ=-X$.

Properties like (\ref{c4}), (\ref{ghz1})-(\ref{ghz4}), predict
perfect correlations between the outcomes of {\em a priori}
uncorrelated measurements on separated particles. In a classical
world, once communication prevented by realizing space-like
separated detections, correlations can arise only if the outcomes
of each measurement on each particle are pre-established. In other
words, a local variable is a list of twelve bits
$\lambda=\{(x_k,y_k,z_k), k=1,2,3,4\}$, where $x_1\in\{-1,+1\}$ is
the pre-established value of a measurement of $X$ on qubit 1, and
so on. The GHZ argument for non-locality aims to show that no list
$\lambda\in\{-1,+1\}^{\times 12}$ can account for all the fifteen
properties above. To verify this, one replaces each Pauli matrix
with the corresponding pre-established value: all the properties
are supposed to hold, but now they are written with ordinary
numbers, whose algebra is commutative. Assuming commutativity, the
multiplication of (\ref{ghz1}), (\ref{ghz2}) and (\ref{ghz3})
gives $z_1y_2x_3y_4=+1$, in contradiction with (\ref{ghz4}) that
reads $z_1y_2x_3y_4=-1$. Therefore, no local variable $\lambda$
can account for all the properties of the list. Of course, a
similar argument could be worked out using others among the
fifteen conditions above.

All in all, by inspection one sees that local variables can
account for thirteen out of the fifteen properties associated to
commuting observables: e.g., using $+1$ as the pre-established
value for all twelve measurement, one fulfills all properties but
(\ref{ghz4}) and $(E_1)\times(E_2)\times(E_3)$ that reads
$YXYZ=-1$. The same is true for the four-qubit GHZ state
$\ket{GHZ_4}$. This rapid argument would suggest that the
non-locality of the cluster state is after all not too different
from the one of the GHZ state. The rest of the paper will show
that the opposite is true.

\subsection{Bell-type inequality}
\label{secb4}

The GHZ argument for non-locality involves identifying properties
that are fulfilled with certainty. Thus, this approach strongly
relies on the details of the state and is not suited for
comparison between different states; nor can it incorporate the
effect of noise in a simple way. Therefore, especially to deal
with experimental results, it is convenient to introduce linear
Bell inequalities.

The best-known inequality for 4-qubit states is the
Mermin-Ardehali-Belinski-Klyshko (MABK) \cite{mabk} inequality
${\cal M}_4$. For the cluster state, after optimizing on the
settings, one finds $\left\langle{\cal M}_4
\right\rangle_{\phi_4}=2\sqrt{2}$ where the local-variable bound
is set at 2. This is indeed a violation, but a rather small one: a
two-qubit singlet attains this amount as well, and $\ket{GHZ_4}$
reaches up to $4\sqrt{2}$. It is well-known by now that the MABK
detects optimally GHZ-type non-locality, but it can be beaten by
other inequalities for other families of states \cite{spectr}.

In our case, it is natural to guess a Bell inequality out of the
GHZ argument: one takes the very same four conditions ,
(\ref{ghz1})-(\ref{ghz4}) that have led to the GHZ argument, and
writes a suitable linear combination of them. Specifically, on the
one hand, the previous results imply that the Bell operator \ba
{\cal B}&=& AIC'D\,+\, AICD'\, +\, A'BCD\,-\, A'BC'D'
\label{bellop}\ea reaches 4 when evaluated on $\ket{\phi_4}$ for
the setting $A=X$, $A'=Z$, $B=Y$, $C=Y$, $C'=X$, $D=Z$ and $D'=Y$.
This is the algebraic value, obviously no state can ever give a
larger value (in particular, $\ket{\phi_4}$ is an eigenstate of
${\cal B}$ for these settings). On the other hand, the classical
polynomial corresponding to ${\cal B}$ satisfies the inequality
\ba \left| ac'd\,+\, acd'\, +\, a'bcd\,-\, a'bc'd'\right| &\leq &
2\,, \label{bi}\ea as one can verify either by direct check, or by
grouping 1 and 2 together, thus recovering the polynomial that
defines the three-party Mermin inequality \cite{mabk,lluis}. The
so-defined four-qubit Bell inequality cannot be formulated using
only four-party correlation coefficients, thus it does not belong
to the restricted set classified by Werner-Wolf and by
\.{Z}ukowski-Brukner \cite{wwzb}. Moreover, on particle 2 only one
non-trivial setting is measured; in other words, no locality
constraint is imposed on it \cite{chine}.

Our inequality exhibits a remarkable feature: the GHZ state
$\ket{GHZ_4}$ does not violate it. The most elegant way to prove
this statement consists in writing down explicitly the projector
$Q$ associated to $\ket{GHZ_4}$: only terms with an even number of
Pauli matrices appear. Consequently,
$\mbox{Tr}\big(Q\,(AIC'D)\big)= \mbox{Tr}\big(Q\,(AICD')\big)=0$
for any choice of the measurement directions, and so
$\mbox{Tr}\big(Q\,{\cal B}\big)= \mbox{Tr}\big(Q\,(A'BCD)\big)-
\mbox{Tr}\big(Q\,(A'BC'D')\big)$ whose algebraic maximum is 2. We
checked also our inequality on the state $\ket{W_4}=\frac{1}{2}
\big(\ket{0001}+\ket{0010}+\ket{0100} + \ket{1000}\big)$ and found
numerically a violation $\moy{{\cal B}}_W\approx 2.618$. In
conclusion, our specific derivation results in a Bell inequality
which acts as a strong entanglement witness for the cluster state
$\ket{\phi_4}$: it is violated maximally by it, and is not
violated at all by the four-qubit GHZ state \cite{note0}.

\section{GHZ argument for the N-qubit state $\ket{\phi_N}$ on a 1-dimensional cluster}
\label{secN1}

The non-locality of the four-qubit cluster state has been studied
in full detail, starting from the expression of $\ket{\phi_4}$.
With the insight gained there, we can move on to look for the
non-locality of the cluster state of an arbitrary number of qubits
$N$, still defined on a one-dimensional lattice. We don't need to
give $\ket{\phi_N}$ explicitly, but can work directly on the set
(\ref{family}) of $N$ eigenvalue equations that define it: \ba \begin{array}{ccl} XZIII...I\,=\,+1&&(E_1)\\ ZXZII...I\,=\,+1&&(E_2)\\
IZXZI...I\,=\,+1&&(E_3)\\ IIZXZ...I\,=\,+1&&(E_4) \\ \vdots\\ II...IZX\,=\,+1&&(E_N)\\
\end{array} \label{eigen} \ea The
GHZ argument appears out of these equations as follows. We focus
on $E_2$, $E_3$ and $E_4$ first. Using the algebra of Pauli
matrices, one derives the
three properties \ba \begin{array}{lcl} C_1=E_2E_3:&&ZYYZI...I\,=\,+1 \\
C_2=E_3E_4:&&IZYYZ...I\,=\,+1\\
C_3=E_2E_3E_4:&&ZYXYZ...I\,=\,-1\,. \end{array} \ea Moving to the
level of local variables (that is, using commutative
multiplication), $E_3C_1C_2$ lead to $z_1y_2x_3y_4z_5=+1$, that
manifestly contradicts $C_3$. The argument is absolutely identical
using $\{E_k,E_{k+1},E_{k+2}\}$, for any $k\in\{2,...,N-3\}$,
because all the eigenvalue equations but the first and the last
one are obtained from one another by translation; and it can be
verified explicitly that it holds also for $k=1$ and $k=N-2$. In
conclusion, we have shown that one can build the following GHZ
argument on five qubits out of any three consecutive eigenvalue
equations:
\begin{itemize}
    \item Take the three equations $\{E_k,E_{k+1},E_{k+2}\}$, for
    $k\in\{1,...,N-2\}$;
    \item With the algebra of Pauli matrices, define
    $C_1=E_kE_{k+1}$, $C_2=E_{k+1}E_{k+2}$ and
    $C_3=E_kE_{k+1}E_{k+2}$, this last property providing the needed
    minus sign;
    \item With commutative algebra, the condition
    obtained as $C_1C_2E_{k+1}$ is exactly the opposite as $C_3$.
\end{itemize}
Let's focus again on the GHZ paradox using $\{E_2,E_3,E_4\}$ for
definiteness: this paradox involves non-trivial operators only on
qubits 1 to 5. This means that one can forget completely about the
other $N-5$ qubits, that is, the partial state $\rho_{12345}$
obtained by tracing out all the other qubits exhibits a GHZ-type
non-locality. This state is certainly {\em mixed} because
$\ket{\phi_N}$ is not separable according to any partition. Since
this is true for any translation, we conclude that any 5-qubit
partial state on consecutive qubits leads to a GHZ argument for
non-locality. The converse holds too: the GHZ argument works only
for {\em consecutive} qubits \cite{note1}. In fact, to obtain the
minus sign that is necessary for the GHZ argument, one has to
multiply three equations that have non-trivial operators on a
common site: a rapid glance to (\ref{eigen}) shows that this can
only be the case if the three equations are consecutive. This GHZ
argument for mixed states reminds the notion of "persistency"
\cite{br1}: one can measure many qubits, or even throw them away,
and strong locality properties are not destroyed. Finally note
that in the GHZ argument involving $\{E_k,E_{k+1},E_{k+2}\}$,
particles $k-1$ and $k+3$ are only asked to measure $Z$: as we saw
for the four-qubit state, on these particles we don't impose any
locality constraint, but they must be asked for cooperation in
order to retrieve the GHZ argument.

Finally note that a Bell inequality can be derived from the GHZ
argument as it was done in Section \ref{sec4q}. On the five
meaningful qubits, the Bell operator reads \ba {\cal
B}&=&(AB)C'(DE)+(A'B')C(DE)\nonumber\\&&+
(AB)C(D'E')-(A'B')C'(D'E') \label{bigen}\ea where we have grouped
the terms in order to make explicit the analogy with Mermin's
inequality \cite{mabk}. The inequality for local variables reads
$|{\cal B}|\leq 2$; partial states of a cluster state violate it
up to the algebraic limit for $A=E=I$, $A'=E'=Z$, $B=D=Z$,
$B'=D'=Y$, $C=Y$ and $C'=X$.

\section{GHZ argument for cluster states on any-dimensional clusters}
\label{secNd}

As a last extension, we consider the non-locality of a cluster
state prepared on two- and more-dimensional square lattices. It is
clear why this problem is not immediately equivalent to the one we
have just studied: the eigenvalue equations (\ref{family}) don't
have the same form as those for one-dimensional lattices
(\ref{eigen}), because the structure of the neighborhood is
different. Consequently, the $N$-qubit cluster state on a
two-dimensional lattice is different from the $N$-qubit cluster
state on a one-dimensional lattice. Still, one expects similar
properties to hold. Indeed, we provide a generalization of the GHZ
argument for cluster states constructed on square lattices of any
dimension.

As a case study, we consider the simplest two-dimensional square
lattice, which is $3\times 3$, because a $2\times 2$ lattice is
equivalent in terms of neighbors to a closed four-sites
one-dimensional loop and we have already solved that case
implicitly \cite{note2}. The nine eigenvalue equations $(E_{ij})$,
$i,j\in\{1,2,3\}$, can be written formally in a way reminiscent of
the lattice: \ba
\begin{array}{ccl} \left(\begin{array}{ccc} X&Z&I\\ Z&I&I\\ I&I&I\end{array}\right)\,=\,+1&&(E_{1,1})\\
\left(\begin{array}{ccc} Z&X&Z\\ I&Z&I\\ I&I&I\end{array}\right)\,=\,+1&&(E_{1,2})\\ \vdots\\
\left(\begin{array}{ccc} I&Z&I\\ Z&X&Z\\
I&Z&I\end{array}\right)\,=\,+1&&(E_{2,2})\\ \vdots\\
\end{array} \label{eigen2} \ea with obvious notations. The basic
reasoning to find the GHZ argument is as before: one can find such
an argument if and only if a minus sign can be produced, that is,
if one takes at least three equations that lead to the product
$ZXZ$ in a site. In this case, the argument is constructed in a
way similar as in the case of one-dimensional lattices. For
instance, if one takes $\{E_{1,1}, E_{1,2} , E_{2,2}\}$, then the
$ZXZ$ product can be found in site $(1,2)$ of the lattice. First,
with the Pauli commutation relations, one builds
$C_1=E_{1,1}E_{1,2}$, $C_2=E_{1,2}E_{2,2}$ and
$C_3=E_{1,1}E_{1,2}E_{2,2}$, where the minus sign appears in
$C_3$. Then, using commutative multiplication, one gets that
$C_1\,C_2\,E_{1,2}$ is exactly the opposite property as obtained
directly from $C_3$. In this example, particles in sites $(3,1)$
and $(3,3)$ can be traced out because in all conditions the
operator in those sites is the identity; and the four particles in
sites $(1,3)$, $(2,1)$, $(2,3)$ and $(3,2)$ undergo a single
measurement ($Z$) and are therefore there to help establishing the
argument.

In general, it is easily checked that: \begin{itemize}
    \item A GHZ argument can be obtained if the three sites form a
    neighbor-to-neighbor path, like $\{E_{1,1}, E_{1,2} ,
    E_{2,2}\}$, or $\{E_{1,1}, E_{1,2} , E_{1,3}\}$; in this case,
    the argument goes as in the examples above.
    \item A GHZ argument cannot be obtained if the three sites do not form a
    neighbor-to-neighbor path, like $\{E_{1,1}, E_{2,2} ,
    E_{3,3}\}$ or $\{E_{1,1}, E_{1,2} ,E_{2,3}\}$.
\end{itemize}
With this characterization, it is obvious how to generalize the
GHZ argument either to larger two- and more-dimensional clusters.
Again, a Bell inequality can be derived from this GHZ argument,
exactly as we did in the previous Sections.

\section{Comparison with graph states}
\label{secgraph}

\subsection{On the extension of our results}

Cluster states are members of a large family of states called {\em
graph states} \cite{hein}. Graph states differ from one another
according to the graph on which the state is built. Since the
definition of the family of commuting operators on the graph is
always (\ref{sa}), our techniques can be applied to study the
non-locality of any graph state. However, the specific results can
be strongly dependent on the graph, which here was the regular
lattice or cluster. For instance, N-qubit GHZ states are graph
states, but --- contrary to what has just been described for
cluster states --- no GHZ argument for their partial states can be
found, because all the partial states are separable. This derives
from the connectivity of the corresponding graph, in which all the
sites are connected only through a single site $a$; therefore, the
operator $S_a$ must be used to find any GHZ argument, and this
operator is non-trivial on all sites.

\subsection{Comparison with systematic inequalities}

Very recently, a systematic way of constructing Bell's
inequalities for any graph state has been found \cite{guehne}. The
result is very elegant: the sum of the elements of the stabilizer
group provides a Bell inequality. It is instructive to apply this
formalism to the four-qubit cluster state, for which we have
provided inequality (\ref{bi}) above. We have written explicitly
the equations corresponding to each element of the stabilizer
group at the beginning of \ref{secghz4}; at then end of the same
paragraph, we have stressed that only fourteen (thirteen
non-trivial plus the identity) of the sixteen equations can be
satisfied in a local variable theory. By definition, QM fulfills
all the properties with the cluster state. Therefore we have a
Bell-type inequality $\widetilde{{\cal B}}_{lv}\leq 14\times
(+1)+2\times (-1)=12$ \cite{note3} for which QM reaches the value
$\widetilde{{\cal B}}_{QM}=16$. This inequality uses three
settings per qubit.

There is a strong link between this inequality and our one
(\ref{bi}). By summing over all the stabilizers, the polynomial
$\widetilde{{\cal B}}$ contains the four terms of (\ref{bellop}),
the four terms which build the symmetric version of it (GHZ
argument based on $YXYZ=-1$), and eight more terms. These
additional terms turn out to be "innocuous" as far as local
variables are concerned: thus, the violation of $\widetilde{{\cal
B}}\leq 12$ is nothing but the simultaneous violation of
(\ref{bi}) and of its symmetric version. However, the two
inequalities are not equivalent on all quantum states, as can be
seen on the GHZ state $\ket{GHZ_4}$ by the same argument as in
\ref{secb4}: eight terms in the polynomial $\widetilde{{\cal B}}$
are products of three Pauli operators, so $\widetilde{{\cal
B}}_{GHZ}\leq 8$ (and the bound can actually be attained). So, for
the inequality discussed in this paragraph, the GHZ state cannot
even reach the local-variable bound.

In summary, in the case of the four-qubit cluster state
$\ket{\phi_4}$, the inequality built according to the recipe of
Ref.~\cite{guehne} exploits the same non-locality as our
inequality (\ref{bi}). Note also that our inequality is easier to
test experimentally because it requires fewer settings (two
instead of three per qubit) and fewer terms (four instead of
fifteen). However, when we apply our method to an arbitrary number
$N$ of qubits (see end of Section \ref{secN1}) we find an
inequality whose violation is always by a factor 2, irrespective
of $N$; whereas the inequalities discussed in Ref.~\cite{guehne}
are such that the violation increases with $N$.

\section{Conclusion}

In conclusion, we have found that a rich non-locality structure
arises from the peculiar, highly useful entanglement of cluster
states of qubits. This non-locality is very different from the one
of the GHZ states: the qualitative difference is most strikingly
revealed by the existence of a GHZ argument for mixed states. In
the 4-qubit case, we have also provided a quantitative witness in
terms of a Bell inequality, that will be an important tool in
planned experiments to produce photonic cluster states
\cite{weinf}.

V.S. and A.A. acknowledge hospitality from the Institut f\"ur
Experimentalphysik in Vienna, where this work was done. We thank
Nicolas Gisin, Barbara Kraus, Sofyan Iblisdir and Lluis Masanes
for insightful comments. This work was supported by the European
Commission (RAMBOQ), the Swiss NCCR "Quantum photonics", the
Austrian Science Foundation (FWF), the Spanish MCYT under "Ram\'on
y Cajal" grant, and the Generalitat de Catalunya. M.A.
acknowledges additional support from the Alexander von Humboldt
foundation.

\end{multicols}


\begin{thebibliography}{10}
\bibitem{note2} We consider the one-dimensional lattice as
an open segment. It turns out by later inspection that the
non-locality properties described in this paper do not change if
the lattice would be a closed loop (that is, if qubits $N$ and 1
were taken to be neighbors). This does not imply that these states
are equivalent.
\bibitem{note1} One can transfer entanglement from
one site to another by suitable measures: this is the basic
insight behind the one-way quantum computer. Here we focus on the
non-locality of the cluster state without any processing.
\bibitem{note3} In a local variable theory, no particular geometric relation
is assumed between $X$, $Y$ and $Z$; so the result holds for
general operators $A_k$, $B_k$ etc., as it should for a Bell-type
inequality.
\end{thebibliography}
\end{document}